# Comparative study of two multiscale thermomechanical models of polycrystalline shape memory alloys: Application to a representative volume element of titanium-niobium

**M.D. Fall[1], E. Patoor[1], O. Hubert[2], K. Lavernhe - Taillard[2]**

[1]Georgia-Tech Lorraine (Georgia-Tech/CNRS UMI 2958), F-57070 Metz, France

[2] LMT (ENS-Paris Saclay/CNRS /Université Paris-Saclay), F-94235 Cachan cedex, France

Abstract: This paper presents a comparative study between two micro-macro modeling approaches to simulate stress-induced martensitic transformation in shape memory alloys (SMA). One model is a crystal plasticity based model and the other describes the evolution of the microstructure with a Boltzmann-type statistical approach. Both models consider a self-consistent scheme to perform the scale transition from the local thermomechanical behavior to the global one. The way the two modeling approaches describe the local behavior is analyzed. Similarities and differences are pointed out. Numerical simulations of the thermo-mechanical behavior of an isotropic titanium-niobium SMA are performed. These alloys have known a growing interest of scientific community given their high potential for application in the biomedical field. Stress-strain curves obtained from the two simulations are compared with experimental results. Evolutions of volume fractions of martensite variants predicted by the two approaches are compared for <100>, <110> and <111> tensile directions. Due to the absence of comparative studies between multiscale models dedicated for SMA, this paper fills a gap in the state of the art in this field and provides a significant step toward the definition of an efficient numerical tool for the analysis of SMA behavior under multiaxial loadings.

## 1. Introduction

Since the 80's the modeling of shape memory alloys (SMA) behavior presents an everlasting interest [Cissé, 2016]. The large field of applications met by these alloys [Mohd Jani, 2014] combined with the relative complexity of their behavior explain this interest. Three main groups of models are classically distinguished. Most part of the modeling effort has been devoted to the development of phenomenological macroscopic approaches. Popularity of this group of models comes from their ability for implementation in finite element softwares for structural analysis [Chemisky, 2011, Aurichio, 2014, Lagoudas, 2012] and for the relatively easy calibration of the material parameters involved [Chemisky, 2015]. Phase field approaches constitute a second group of models aiming at the description of the microstructure evolution during any thermomechanical process.

Several examples of these models can be found in [Mamivand, 2013]. The third group of models aims at determining the macroscopic behavior of the materials through a description of its microstructure and a modeling of local strain mechanisms [Lagoudas, 2006]. The present paper focuses on this third group called multiscale modeling. The objective of the present work is to investigate multiaxial aspects of the mechanical behavior of Ti-Nb alloys as a new class of shape memory alloys [Laheurte, 2010, Miyazaki, 2006]. This is performed from a numerical point of view considering the lack of commercial Ti-Nb alloys products (plates, tubes, sheets, …) required for a complete multiaxial experimental characterization. Multiscale modeling can then be considered as a virtual testing machine for material evaluation. The choice of a particular multiscale model is not an easy task. Many multiscale models have been developed and published in the literature but to the authors' best knowledge, the capabilities of these various models have never been compared. An important Round-robin for SMA modeling was performed within the ESF S3T EUROCORE project [Sittner, 2009] but only phenomenological macroscopic approaches have been considered. In an attempt to fill this gap, two different multiscale approaches are considered in the present work and applied to the modeling of a TiNb alloy superelastic behavior.

## 2. Multiscale modeling of SMA

Many multiscale models, based on micromechanics, are used to describe the quasi-static behavior of SMAs. Three characteristic microstructural scales are classically considered in these models (Figure 1):

- The phase: either austenite or martensite variants. Each phase is characterized by its crystallographic structure and its properties (resistivity, stiffness, entropy, transformation strain ...)
- The grain or crystal: according to the evolution and direction of transformation, the grain may consist in single phased austenite, martensite variants or in a mixture of both phases in separated domains. The local strain mechanism is described at this length scale.
- The polycrystal: representative of an aggregate of grains with given orientations separated by grain boundaries. The crystallographic texture of parent phase is the relevant information at this length scale.

Models mainly differ in the way the crystal behavior is modeled. We mainly distinguish two different approaches:

- Plasticity based models: these models are inspired by crystal plasticity models of metallic alloys (steels, nickel, or cooper-based alloys). A like-Schmid law is considered and evolution of the transformation is expressed thanks to a consistency condition. Most of the models describing the SMA behavior belong to this category. Some models consider a single martensite variant with the assumption that only one main variant may appear during a superelastic monotonic uniaxial loading. The multivariant/multidomain modeling approaches mainly differ with each other regarding the definition of the intragranular internal stresses. The latter can be expressed as a constant value, as a full interaction matrix or as a simplified matrix composed of two independent terms associated with compatible and incompatible domains respectively. It can also be estimated from a second scale transition scheme in which domains are seen as individual oblate inclusions. The interesting reviews from [Cissé, 2016, Lagoudas, 2006] can be referred to for more details.

- Statistical models: Less usual, this modeling expresses the microstructure as a statistical distribution. Statistical functions are used to estimate the evolution of the transformation [Fischlschweiger, 2012].

We can also classify models according to the adopted scale transition schemes to link grain scale and polycrystalline scale. Mori-Tanaka scheme, self-consistent one, uniform stress/strain approaches or finite elements based analyses may be used to perform the scale transition.

An important drawback for multiscale models is their high computational cost depending on the complexity of the local description. The local stress field is strongly multiaxial, even for a macroscopic simple tension, due to grain to grain interactions or to variant selection which may be activated or deactivated at each loading step. Our motivation in this study is to compare, for the first time in the framework of shape memory alloys behavior, the formulations and performances of two different multiscale approaches.

In a first part, we introduce the micromechanical model proposed by [Siredey, 1999] which describes the local thermomechanical behavior inside a single grain considering the crystallographic nature of the martensitic transformation. Volume fractions of martensite variants are chosen as internal variables and their evolution is derived from the definition of a thermodynamic potential at the grain scale. In a second part, we present the micromechanical model proposed by [Maynadier, 2011], which describes the behavior of a representative volume of polycrystalline SMA where a Boltzmann type statistical law allows the volume fractions of martensite variants and of austenite to be calculated. Finally, the two models are used to simulate the superelastic behavior of a titanium-niobium (TiNb) alloy.

TiNb alloys are good candidates for biomedical applications. We can cite the study from [McMahon, 2012] where TiNb alloys exhibit reduced ion release and better corrosion resistance, compared to NiTi alloys. Some TiNb alloys exhibiting a very low Young modulus, close to bone stiffness (which is between 10 and 30 GPa), are presented in [Elmay, 2013, Hon, 2003]. These TiNb alloys are especially good candidates for manufacturing bone implants. They reduce both the cytotoxicity and stress-shielding, which defines bone loss mainly observed with high stiffness implants [Piotrowski, 2012].

Unfortunately, few experimental mechanical characterizations of these alloys are nowadays available. Most of these experimental data comes from tensile loadings since TiNb are most of the time available in a wire form. Relevant multiscale modeling of TiNb alloys would allow a first estimation of the multiaxial behavior of these alloys.

# 3. [Siredey, 1999] and [Maynadier, 2011] Multiscale models

## 3.1 Single crystal model of [Siredey, 1999]

This 3D multivariant model relies on micromechanics to propose a simplified expression of the interaction energy between the martensite variants in the material. An interaction matrix $H^{nm}$ is used for the description of the interactions between martensitic variants $n$ and $m$ with respective volume fractions $f^n$ and $f^m$. In the framework of small perturbations, total strain $\varepsilon_n$ of a variant $n$ of volume $V_n$ is the sum of elastic and transformation components Eq(1):

$$\boldsymbol{\varepsilon}_n = \boldsymbol{\varepsilon}_n^{el} + \boldsymbol{\varepsilon}_n^{tr} \tag{1}$$

Habit planes between austenite and martensite variants are defined by the unit normal to the habit plane $\vec{n}$ and the unit direction of transformation $\vec{m}$ with an amplitude $g$. The transformation strain $\epsilon_n^{tr}$ of each $n$ variant is:

$$\boldsymbol{\varepsilon}_n^{tr} = \frac{1}{2} g(\vec{n} \otimes \vec{m} + \vec{m} \otimes \vec{n}) \tag{2}$$

Free energy for the whole grain is expressed in Eq(3).

$$\Psi = -B(T - T_0) \sum_n f^n + \boldsymbol{\sigma}_g \sum_n \boldsymbol{\varepsilon}_n^{tr} f^n + \frac{1}{2} \boldsymbol{\sigma}_g : \mathbb{S} : \boldsymbol{\sigma}_g - \frac{1}{2} \sum_{n,m} \boldsymbol{H}^{nm} f^n f^m \tag{3}$$

$\sigma_g$ and $T$ are the applied stress and temperature at the grain scale, $B$, $\mathbb{S}$, and $T_0$ are respectively the sensitivity parameter to chemical energy evolution, the compliance tensor and the reference temperature.

The transformation starts when the thermodynamic force $F^n = \frac{\partial \Psi}{\partial f^n}$ associated with the internal variable $f^n$ reaches a critical value $\pm F^C$ that is characteristic of the material.

$$df^n = \sum_m [\boldsymbol{H}^{nm}]^{-1} (\boldsymbol{\varepsilon}_n^{tr} d\boldsymbol{\sigma}_g - BdT) \tag{4}$$

The evolution of fraction $df^n$ in Eq(4) is obtained from energy derivation and application of consistency condition ($dF^n = 0$).

## 3.2 Single crystal model: [Maynadier, 2011]

The model is based on the comparison of the Gibbs free energy densities of each martensite variant *n* and austenite phase *a*[1]. The same hypothesis of homogeneous stress $\sigma_g$ than for Siredey's model is applied at the variant scale, resulting in Gibbs free energy density expression reported in Eq(5) where index *i* indicates *n* variants + *a* phase. The transformation strain is the same as in Eq(2) except for austenite whose transformation deformation is null (as reference deformation).

$$\psi_i = h_i - Ts_i - \boldsymbol{\sigma}_g : \boldsymbol{\epsilon}_i^{tr} - \frac{1}{2}\boldsymbol{\sigma}_g : \mathbb{S} : \boldsymbol{\sigma}_g \qquad (5)$$

$h_i$, $s_i$ and $\mathbb{S}$ are enthalpy density, entropy density, and compliance tensor of the martensite variant or austenite phase. $\sigma_g$ denotes the stress at the grain scale.

A probabilistic estimation of each variant or austenite phase (denoted as variant *n+1*) volume fraction is made using a Boltzmann distribution (see Eq(6)). Interactions at the interfaces are not taken into account. The modeling uses one numerical parameter, $A_s$, which drives interfacial effects. This parameter can be related to the Boltzmann constant and temperature via a statistical volume.

$$f_i = \frac{exp(-A_s\psi_i)}{\sum_{i=1}^{n+1} exp(-A_s\psi_i)} \qquad (6)$$

This formulation allows the term $\frac{1}{2}\boldsymbol{\sigma}_g\mathbb{S}\boldsymbol{\sigma}_g$ in Eq(5) to be removed since it does not change from one variant to another.

In this approach, the microstructure is defined as a distribution and fractions are obtained by direct comparison between Gibbs free energy density levels of the constituents. This strategy is completely different from the strategy used by [Siredey, 1999]. The latter is a threshold model with a fixed critical value for martensite nucleation and an evolution of the fractions derived from the consistency condition.

This model of [Maynadier, 2011] has been recently extended to chemo-magneto-mechanical couplings in magnetic shape memory alloys accounting for thermal exchanges [Fall, 2016].

We can point out similarities between the free energy expressions from Siredey and Maynadier.

[1] The so-called R-phase of equi-atomic NiTi can be considered in this modeling in addition to martensite and austenite phases.

The Gibbs free energy density in Eq(5) can be written at the grain scale:

$$\Psi = \sum_i f^i \Psi_i = \sum_n f^n \Psi_n + (1 - \sum_n f^n)\Psi_a = \Psi_a + \sum_n f^n(\Psi_n - \Psi_a) \quad (7)$$

With $f^n$ the volume fraction of martensite variant *n*, and $\psi_a$ the Gibbs free energy density of austenite phase.

Since stress and compliance are considered homogeneous, one gets from Eq(7):

$$\Psi(\boldsymbol{\sigma_g}, T) = \Psi_a - \frac{1}{2}\boldsymbol{\sigma_g} : \mathbb{S} : \boldsymbol{\sigma_g} + \sum_n f^n(h_n - h_a) - T\sum_n f^n(s_n - s_a) - \boldsymbol{\sigma_g} : \sum_n f^n \boldsymbol{\epsilon}_n^{tr} \quad (8)$$

The energy is known for a constant. We fix: $\Psi(\boldsymbol{\sigma_g} = \mathbf{0}, T = T_0) = 0$ where $T_0$ is a reference temperature. We get so: $\Psi_a + \sum_n f^n(h_n - h_a) = T_0 \sum_n f^n(s_n - s_a)$

This can be introduced in the Gibbs free energy density in Eq(8):

$$\Psi(\boldsymbol{\sigma_g}, T) = \sum_n f^n(s_n - s_a)(T_0 - T) - \frac{1}{2}\boldsymbol{\sigma_g} : \mathbb{S} : \boldsymbol{\sigma_g} - \boldsymbol{\sigma_g} : \sum_n f^n \boldsymbol{\epsilon}_n^{tr} \quad (9)$$

All martensite variants are assumed to exhibit the same entropy: $s_n = s_{ma}$

To get from Eq(9) the Siredey expression in Eq(3), we have to consider:

- B constant is introduced corresponding to the variation of entropy density between austenite and martensite : B=$\Delta s = s_a - s_{ma} \geq 0$
- The interaction between variants that increases the free energy density: $\Psi_{inter} = \sum_{n,m} f^n f^m \mathbf{H}^{nm}$
- a Legendre Transformation (or complementary energy expression): $\psi + \Psi + \Psi_{inter} = \psi(\mathbf{0}, T_0) = 0$

so that : $\Psi = -\psi - \Psi_{\text{inter}}$

It is interesting to notice that the Gibbs free energy expression in Eq(5) concerns each variant in the volume of the grain while expression in Eq(3) is the Helmholtz expression for the whole grain (possibly a mix of austenite and martensite).

## 3.3. Scale transition rules: from grain to polycrystal

Both models are based on thermodynamics and involve transitions from variant to grain then from grain to polycrystalline scale. The macroscopic behavior of the polycrystalline SMA is estimated by averaging the behavior of single grains using the self-consistent scale transition scheme. The latter is relevant to describe aggregates of crystals. The effective tensor $\mathsf{C}_{eff}$ is expressed as: $\mathsf{C}_{eff} = < \mathsf{C}_g : [(\mathsf{C}_g + \mathsf{C}^*)^{-1} : (\mathsf{C}_{eff} + \mathsf{C}^*)] >$ with the Hill constraint tensor $\mathsf{C}^*$. This implicit expression needs a numerical resolution.

## 4. Application to titanium - niobium polycrystalline alloy

Titanium - niobium alloys undergo a cubic ($a_0$=0. 328$nm$) to orthorhombic phase transition (a = 0.318$nm$; b=0.4818$nm$; c=0.464$nm$). The values of cell parameters are taken from measurements made by [Elmay, 2013]. We focus on $TiNb_{26\%at.}$ composition, which leads to a martensitic structure at ambient temperature ($M_s$=265$K$ and $A_f$ =296$K$). The representative volume element (RVE) is defined by a set of 100grains with random orientations to obtain an isotropic crystallographic texture. Homogeneous isotropic elasticity such as $E_{aust.} = E_{mart.} = E = 22$ GPa and $\nu = 0.33$ is assumed.

Parameters used in both modeling are listed in Table 1.

Crystallographic theory of martensite is used to estimate the possible interfaces. This theory is based on the assumption of an invariant plane separating austenite and martensite phases. Based on this theory, we found 2x6 (12) austenite / single martensite habit planes variants (Table 2) and 24 habit planes between austenite and twinned martensite pairs. The 2x6 habit planes variants are denoted $V1^+$, $V2^+$, $V3^+$, $V4^+$, $V5^+$, $V6^+$ and $V1^-$, $V2^-$, $V3^-$, $V4^-$, $V5^-$, $V6^-$. Twinned pairs are formed between variant $I$ and variant $J$ with respective size ratios λ and (1-λ). In our calculations, λ value is close to 1 (λ=0.999), which means that configuration is close to a single variant one. Indeed, simulations considering either single variants or twinned variants give similar results. We only present the results considering the set of 12 possible austenite/single martensite variants in Table 2.

$H^{nm}$ is simplified considering only two terms $H^1$=40MPa and $H^2$=400MPa respectively for compatible and incompatible combinations (Table 3). $B$ value is identified from tensile test measurements: $B = 0.08$ MPa/K.

$A_s = 1.4.10^{-5} m^3.J^{-1}$ is obtained from a differential scanning calorimetry (DSC) measurement (the identification process is explained in [Fall, 2016]).

Figure 2 illustrates a comparison between both modeling during superelastic tensile loading, considering the set of 12 possible austenite / single martensite variants. The simulations are compared with experimental results from [Kim, 2006]. Both models lead to a very similar global stress-strain curve (stress gap less than 4 MPa) until a strain value $E^{11}$=1% ($f^{martensite}$ around 13%). Above this value, the gap between the two curves remains low but increases with the loading: for $f^{martensite}$ =50%, the stress gap is 14MPa; for $f^{martensite}$ =90%, the gap reaches 23MPa.

One advantage of the micromechanical models over macroscopic phenomenological ones is that they give access to the local quantities evolution during the loading process.

We choose to plot in Figure 3 the local stress-strain behavior of some specific grains inside the RVE denoted grain22, grain17 and grain2 where stress axis is close to $< 111 >$, $< 001 >$ and $< 011 >$ crystallographic directions respectively. Both modeling give similar phase transformation kinetic (Figure 3). The values of local stresses are however smaller for [Maynadier, 2011]. The evolution of variants volume fractions as function of stress is plotted in Figure (4) for the three grains. The variants selection for both approaches is very similar especially regarding the *main* variants. More variants are selected in [Maynadier, 2011]. In fact, if we look at grain17, couples of variants ($V1^{+}$, V1), ($V2^{+}$, $V2^{-}$), ($V5^{+}$, $V5^{-}$) and ($V6^{+}$, $V6^{-}$) are selected while only variants $V1^{-}$, $V2^{+}$, $V5^{-}$, $V6^{+}$ nucleate in [Siredey, 1999]. These couples of variants have very close thresholds and lead to the same strain levels. But, in [Siredey, 1999], the hardening effect leads to a unique selection of the best oriented pair of variants.

Figure 4 shows that the local stresses level in each variant is lower for [Maynadier, 2011] than for [Siredey, 1999]. Indeed, the incompatibilities due to heterogeneous variants' selection are not accounted for in this approach. We can notice that the local stresses remain in the same magnitudes for both models (very close values are obtained for grain17 and grain22). Grains with orientations close to $< 011 >$ are the first to transform. Grain2 belongs to this category. In addition, we also made a comparison of the transformation thresholds under biaxial stress condition. Results are reported in figure 5. The shape of the transformation surfaces predicted by both models is very similar..

All these similarities between the numerical predictions provided by both models can be understood considering that the two formulations differ mainly by the interaction term. As a Gibbs formulation, Eq(5) ensures a minimum energy principle and can consequently be used in a Boltzmann probability function. This formulation allows an expression of incremental martensite fraction to be derived at the threshold (Eq(10)), very close to the estimation given in Eq(4) from Siredey's approach, showing that $A_s$ parameter can be related to terms of $\boldsymbol{H}^{nm}$ matrix.

$$df^n = A_s(\boldsymbol{\varepsilon}_n^{tr} : d\boldsymbol{\sigma}_g - BdT) \qquad (10)$$

Consequently [Siredey, 1999] and [Maynadier, 2011] models are expected to give comparable results if parameters are appropriately identified.

5. Discussion

The way the microstructure of the grain is described by both models is very different: distribution of variants without interfaces for [Maynadier, 2011], martensite domains separated by interfaces and presenting invariant habit planes with austenite for [Siredey, 1999].

As expected, the model from [Maynadier, 2011] presents a smaller stress-strain slope in Figure 2 due to homogeneous stress assumption at the grain scale. Indeed, when we look at the behavior of individual variants in Figure 3, we can notice that slopes are lower for [Maynadier, 2011]. This result is consistent. As expected, the appearance of new variants inside a grain does not lead to any hardening effect.

The selection of incompatible variants leads to higher levels of internal stresses for [Siredey, 1999] (through $H^2$ value from interaction matrix). This fact explains the higher slopes in the stress-strain variants behavior (Figure 4). The distribution of volume fractions of variants is such that transformation strains at the grain scale are very close from one modeling to another (Figure 3). Local stresses at the grain scale are the same for both models due to homogeneous stress hypothesis.

6. Limitations of the models

The main limitation of Siredey model is its inability to predict shape memory effects. Indeed, concomitant appearance of all martensite variants during a cooling leads to unrealistic values of internal stresses. DSC curves cannot be modeled properly. On the contrary, [Maynadier, 2011] is able to predict a DSC curve as shown in Figure 6. [Siredey, 1999] is limited for martensite nucleation and reorientation aspects during a mechanical loading.

[Maynadier, 2011] is able to predict shape memory effects too. We illustrate the case of Grain 17 pre-loaded at T=-20°C<Af in Figure 7. The martensite variants are accommodated at the initial stage exhibiting equivalent volume fractions ($f^n$=1/12). A mechanical loading is applied next. During this loading, the best oriented variants are selected among the others whose volume fraction decreases.

Above all, [Maynadier, 2011] presents limitations linked to the reversible framework of the formulation. The introduction of an additional constant Lg in the energy expression is a possible solution to model the mechanical (and DSC) hysteresis (see Fig. 6). Enthalpy density is expressed considering an extra constant +Lg or –Lg respectively for direct and reverse transformation. This simplified formulation of hysteresis, is only relevant if initial and final stages are single phased (100% austenite or 100% martensite). Hence, [Siredey, 1999] appears more accurate for non-monotonic and non proportional loadings. An example of cyclic loading applied to Grain17 is shown in figure 8. With [Siredey, 1999], we can perform inner loops because the effect of loading history is considered. When we try to perform the inner loop with [Maynadier, 2011], the curve directly joints the major cycle. There is no history effect. Moreover a given stress state always leads to the same distribution of variants, no matter the loading path.

## 7. Conclusion

In this work, we performed comparisons between two multiscale models: [Siredey, 1999] proposed a plasticity-based model, and [Maynadier, 2011] a statistics based one. We used crystallographic theory of martensite for calculation of possible habit planes and identification of groups of compatible and incompatible variants for TiNb.

Despite some strong differences highlighted in section 5, both modelings lead to similar martensitic transformation kinetics and similar selection of variants for both uniaxial and biaxial loadings. The local mechanical behavior (considering isothermal conditions) of some specific grains predicted by both models give similar results. The transformation surfaces for proportional biaxial tension - compression loadings are also similar. Similar results have been obtained when both approaches have been applied to another titanium-niobium alloy with a different composition ($TiNb_{24\%at.}$). Moreover models have been tested in anisothermal conditions (accounting for heat exchanges – not presented in the paper) leading to a super-elastic behavior very close to each other.

Considering computational aspects, [Siredey, 1999] uses 30 input parameters and is implemented in C++ language while [Maynadier, 2011], in MATLAB, needs 29 input parameters. The calculation times are quite similar (~10 minutes for a loading until 250MPa with step=0.01MPa) but it is hard to draw a conclusion from this information because the way the two models are implemented is very different.

**Acknowledgment**: The authors gratefully acknowledge the financial support of the Conseil Régional du Grand Est, France.

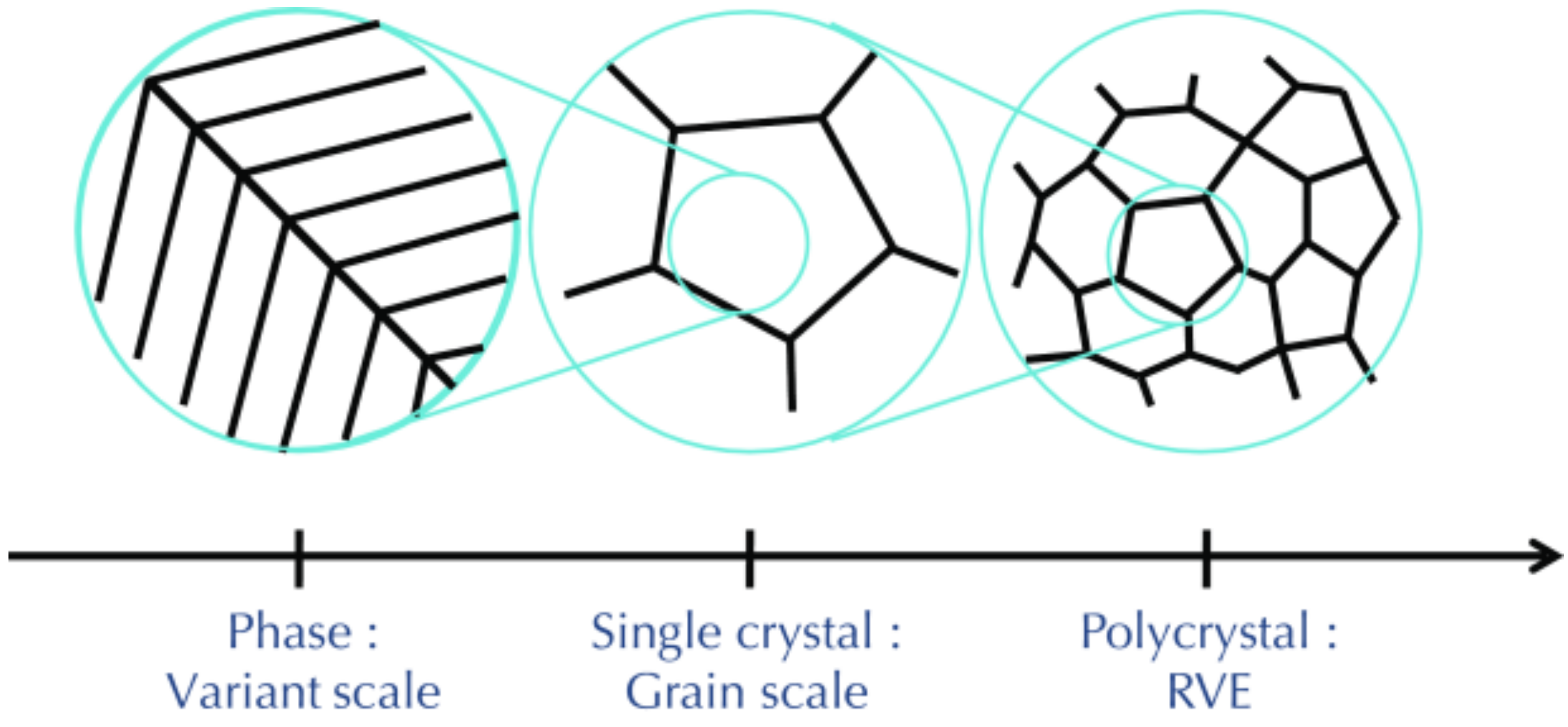

**Figure 1** Characteristic microstructural scales considered in multiscale models

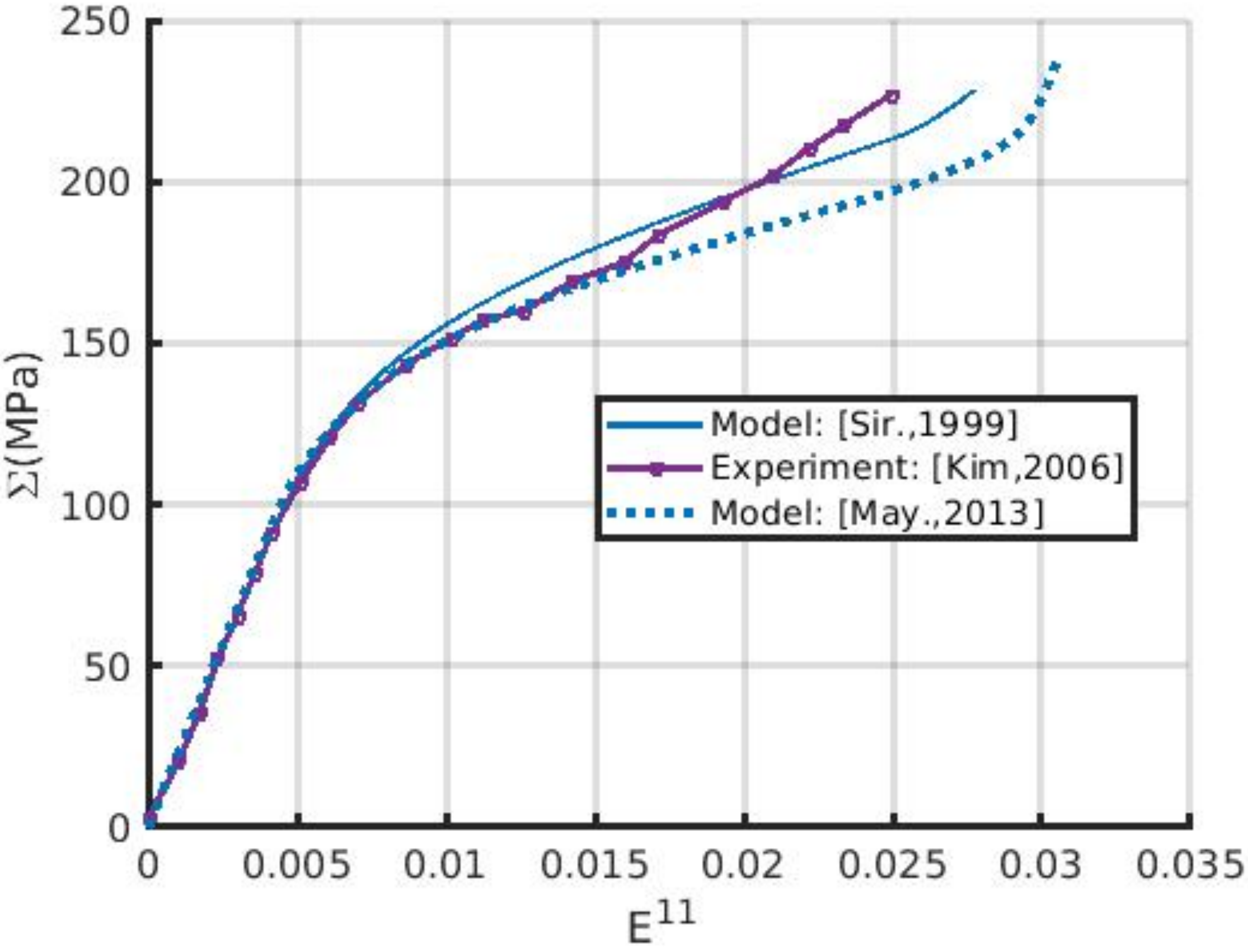

**Figure 2** Simulation of superelastic tensile behavior (T=300K > $A_f$) for $TiNb_{26\%at.}$ Alloys

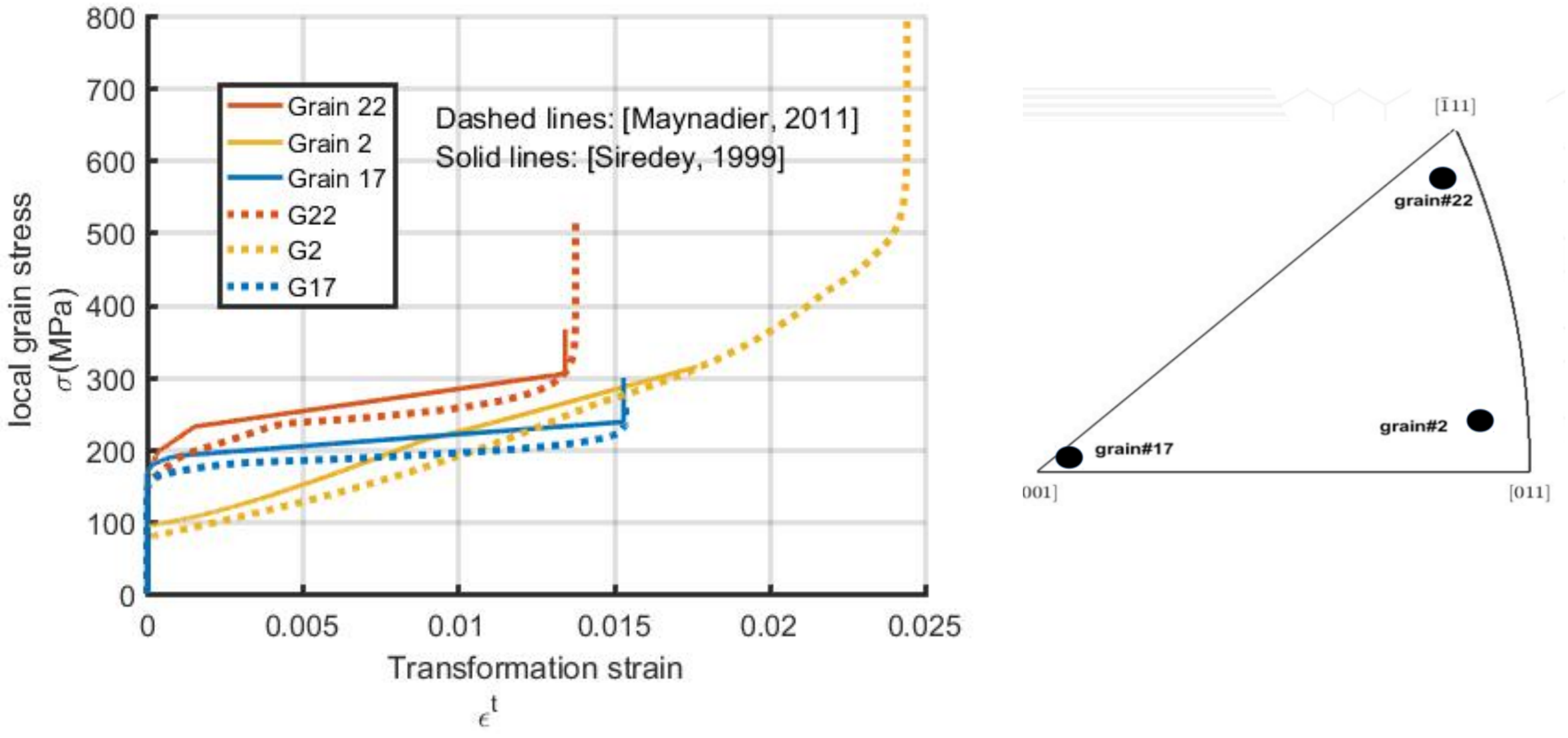

**Figure 3** Local behaviors of specific grains inside the volume

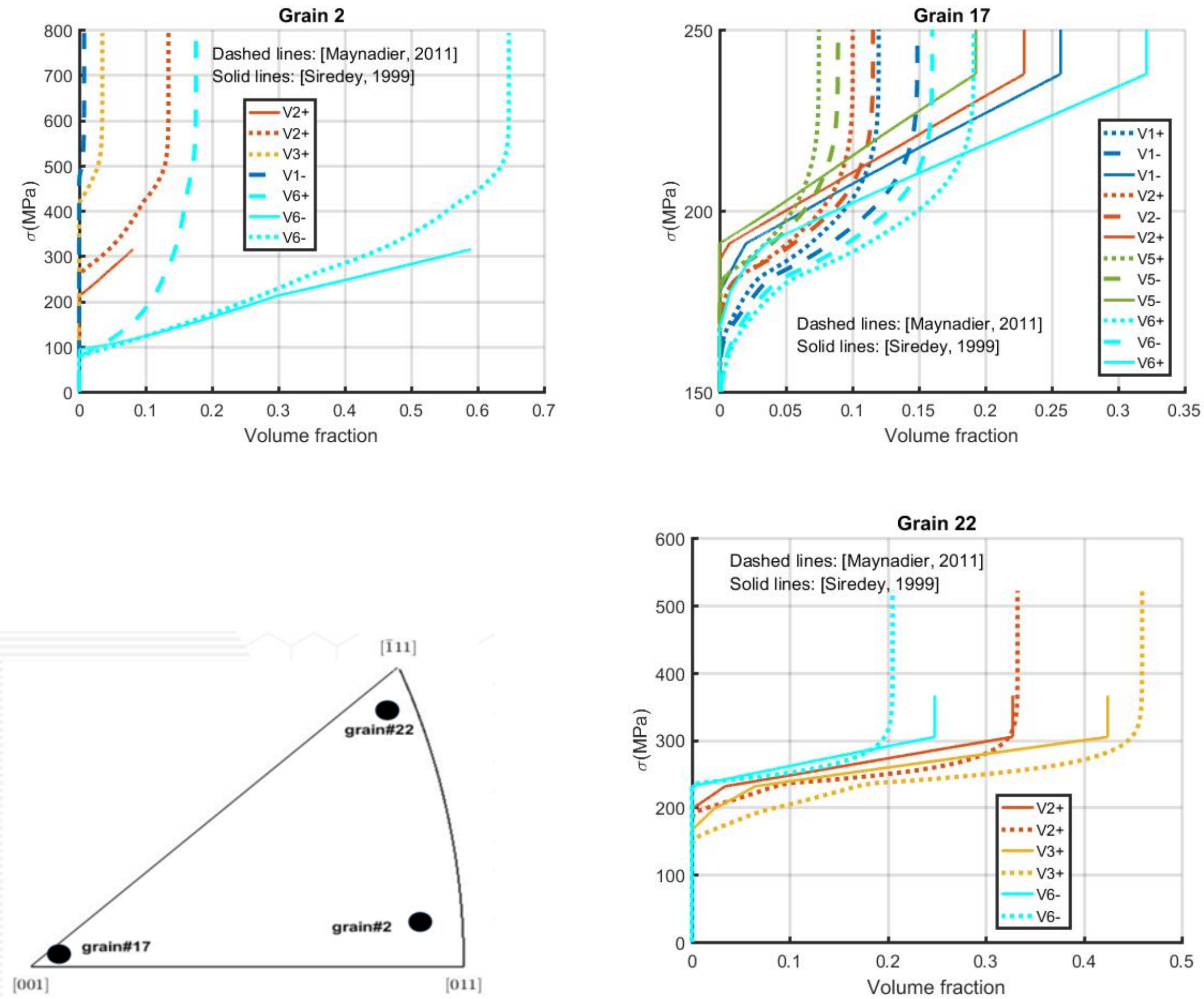

**Figure 4** Selection of variants for specific grains inside thevolume

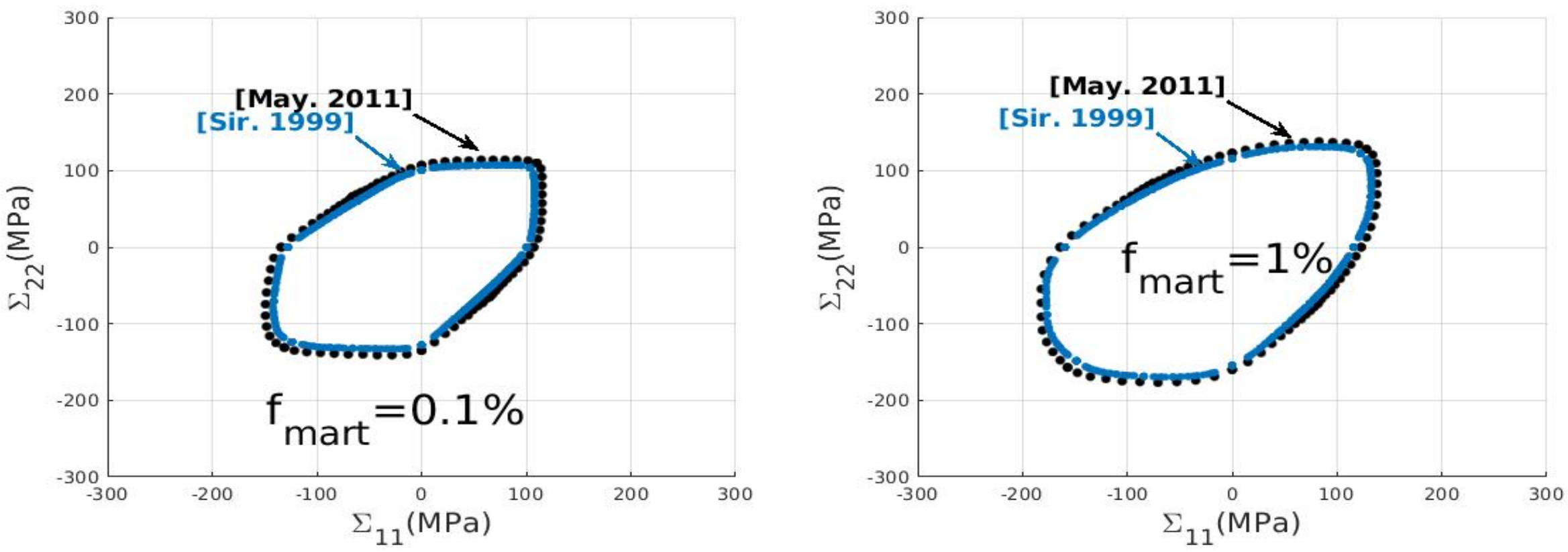

**Figure 5** Biaxial transformation surfaces

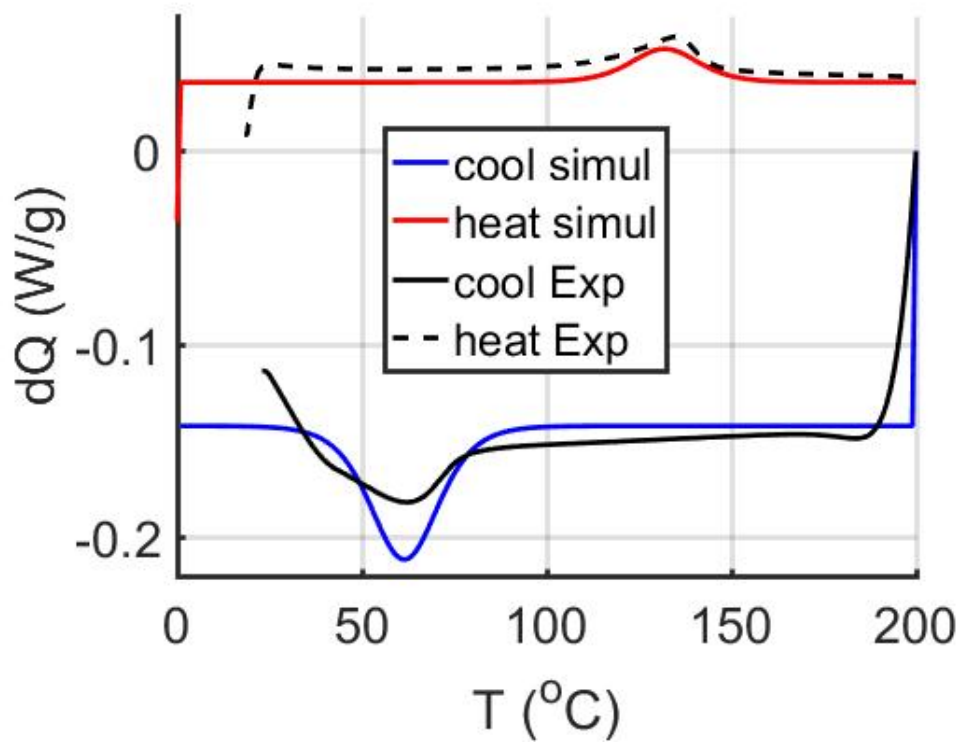

**Figure 6** Experimental and simulated DSC for $TiNb_{24\%at.}$ : $\Delta s$=0.11MPa/K, $T_0$=392.5K, Lg=0.29E+6 $J.m^3$

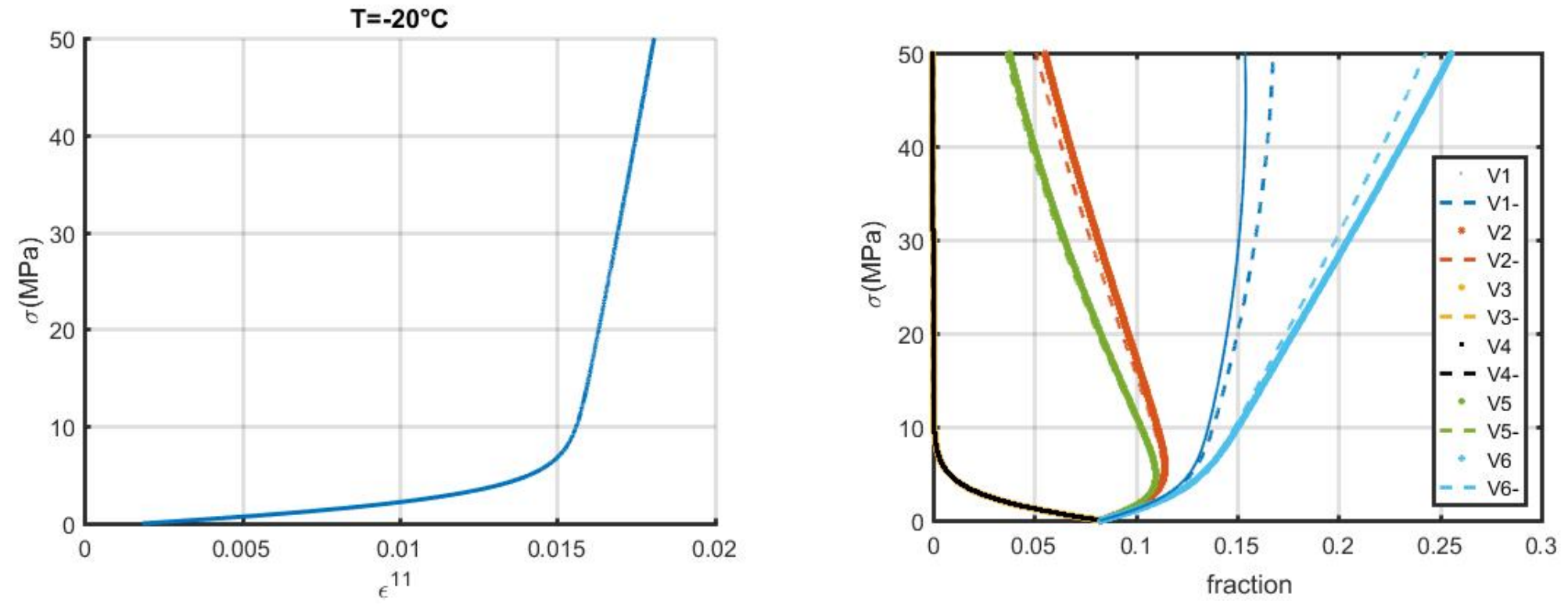

**Figure 7** Reorientation process in Grain 17 (close to <001> direction) during an uniaxial loading at T=-20°C

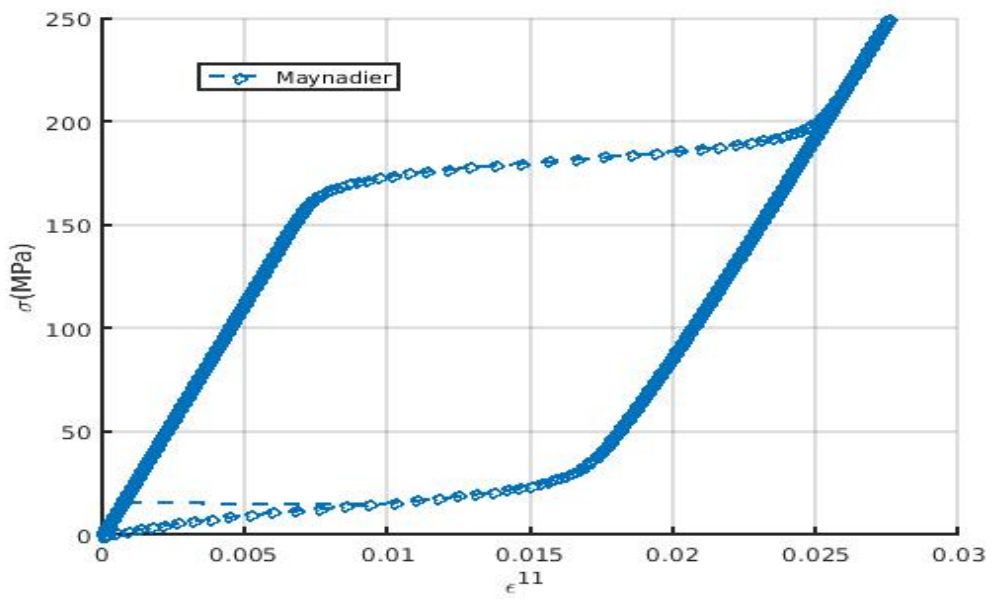

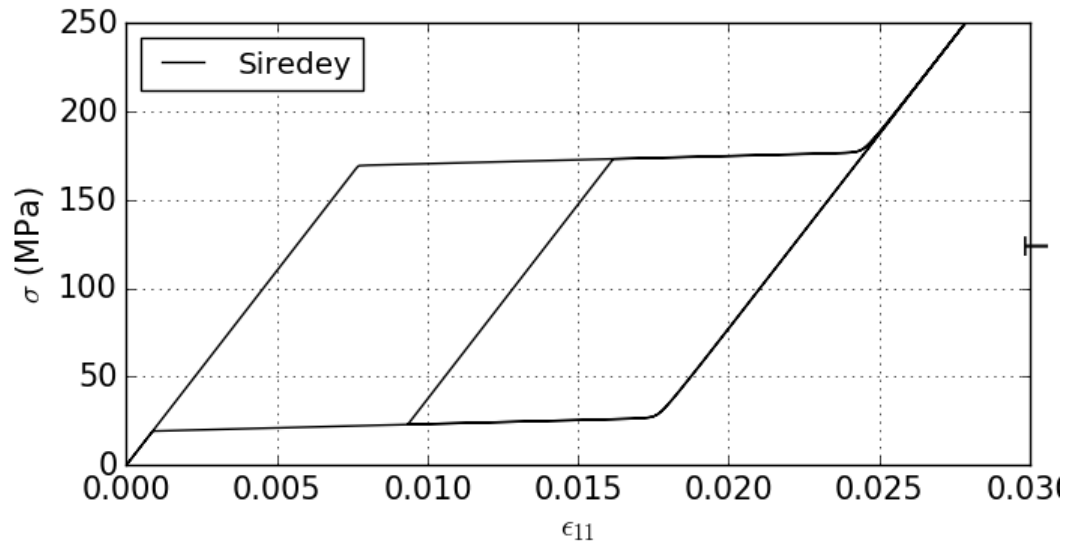

**Figure 8** Cycling tests on Grain 17 (close to <001> direction) at T=27°C (loading steps including a minor cycle: $\sigma_{11\ (MPa)}=$ [0→250 250→15 15→250 250→0]

| [Siredey, 1999] | [Maynadier, 2011] |
|---|---|
| Transformation temperatures: $M_s$=265K, $A_f$=296K<br>Cell parameters: $a_0$= 0.328nm; $a$ = 0.318nm; $b$=0.4818nm; $c$=0.464nm<br>Elastic constants: $E$=22GPa, $\nu$=0.33<br>Habit planes 12x(n,m,g)-→Table 2 | |
| Interaction $H^1$=40MPa, $H^2$=400MPa<br>$B = s_a - s_{ma} = 0.08\ MPa/K$<br>$T_0 = \frac{1}{2}(M_s + A_f)$ | $A_s = 1.4.10^{-5} m^3.J^{-1}$<br>Entropy $\Delta s = s_a - s_{ma} = 0.08\ MPa/K$<br>$\Delta h = h_a - h_{ma} = \Delta s * T_0$ |

Table 1 List of modeling parameters

| | n1 | n2 | n3 | m1 | m2 | m3 |
|---|---|---|---|---|---|---|
| V1+ | -0.658 | 0.533 | 0.533 | 0.683 | 0.517 | 0.517 |
| V2+ | -0.658 | -0.533 | 0.533 | 0.683 | -0.517 | 0.517 |
| V3+ | 0.533 | 0.658 | 0.533 | 0.517 | -0.683 | 0.517 |
| V4+ | -0.533 | -0.658 | 0.533 | -0.517 | 0.683 | 0.517 |
| V5+ | 0.533 | 0.533 | -0.658 | 0.517 | 0.517 | 0.683 |
| V6+ | -0.533 | 0.533 | -0.658 | -0.517 | 0.517 | 0.683 |
| V1- | -0.658 | -0.533 | -0.533 | 0.683 | -0.517 | -0.517 |
| V2- | -0.658 | 0.533 | -0.533 | 0.683 | 0.517 | -0.517 |
| V3- | -0.533 | 0.658 | -0.533 | -0.517 | -0.683 | -0.517 |
| V4- | 0.533 | -0.658 | -0.533 | 0.517 | 0.683 | -0.517 |
| V5- | -0.533 | -0.533 | -0.658 | -0.517 | -0.517 | 0.683 |
| V6- | 0.533 | -0.533 | -0.658 | 0.517 | -0.517 | 0.683 |

**Table 2** Possible austenite/single martensite variant interfaces ($g = 0.055$)

| | V1+ | V2+ | V3+ | V4+ | V5+ | V6+ | V1- | V2- | V3- | V4- | V5- | V6- |
|---|---|---|---|---|---|---|---|---|---|---|---|---|
| **V1+** | ***C*** | ***C*** | ***I*** | ***I*** | ***I*** | ***I*** | ***I*** | ***C*** | ***I*** | ***I*** | ***I*** | ***I*** |
| **V2+** | ***C*** | ***C*** | ***I*** | ***I*** | ***I*** | ***I*** | ***C*** | ***I*** | ***I*** | ***I*** | ***I*** | ***I*** |
| **V3+** | ***I*** | ***I*** | ***C*** | ***C*** | ***I*** | ***I*** | ***I*** | ***I*** | ***I*** | ***C*** | ***I*** | ***I*** |
| **V4+** | ***I*** | ***I*** | ***C*** | ***C*** | ***I*** | ***I*** | ***I*** | ***I*** | ***C*** | ***I*** | ***I*** | ***I*** |
| **V5+** | ***I*** | ***I*** | ***I*** | ***I*** | ***C*** | ***C*** | ***I*** | ***I*** | ***I*** | ***I*** | ***I*** | ***C*** |
| **V6+** | ***I*** | ***I*** | ***I*** | ***I*** | ***C*** | ***C*** | ***I*** | ***I*** | ***I*** | ***I*** | ***C*** | ***I*** |
| **V1-** | ***I*** | ***C*** | ***I*** | ***I*** | ***I*** | ***I*** | ***C*** | ***C*** | ***I*** | ***I*** | ***I*** | ***I*** |
| **V2-** | ***C*** | ***I*** | ***I*** | ***I*** | ***I*** | ***I*** | ***C*** | ***C*** | ***I*** | ***I*** | ***I*** | ***I*** |
| **V3-** | ***I*** | ***I*** | ***I*** | ***C*** | ***I*** | ***I*** | ***I*** | ***I*** | ***C*** | ***C*** | ***I*** | ***I*** |
| **V4-** | ***I*** | ***I*** | ***C*** | ***I*** | ***I*** | ***I*** | ***I*** | ***I*** | ***C*** | ***C*** | ***I*** | ***I*** |
| **V5-** | ***I*** | ***I*** | ***I*** | ***I*** | ***I*** | ***C*** | ***I*** | ***I*** | ***I*** | ***I*** | ***C*** | ***C*** |
| **V6-** | ***I*** | ***I*** | ***I*** | ***I*** | ***C*** | ***I*** | ***I*** | ***I*** | ***I*** | ***I*** | ***C*** | ***C*** |

**Table 3** Shape of the interaction matrix (C=Compatible I=Incompatible)